\documentclass[epj,referee]{svjour}
\usepackage{amssymb,color,graphicx,epsfig}
\usepackage{dcolumn}
\usepackage{bm}

\input{tcilatex}

\begin{document}

\title{The effect of clusterings on the equilibrium states of local
majority-rule: Occurrence probability and robustness}
\author{Yu-Pin Luo\inst{1} \, Jinn-Wen Wu\inst{2} \and Ming-Chang Huang%
\inst{3}\thanks{\emph{email:} mchuang@cycu.edu.tw}}
\institute{Department of Electronic Engineering, National Formosa University, Huwei,
63201 Taiwan \and Department of Applied Mathematics, Chung-Yuan Christian
University, Chungli, 32023 Taiwan \and Department of Physics, Chung-Yuan
Christian University, Chungli, 32023 Taiwan}
\date{Received: date / Revised version: date}
\maketitle

%
%

%
\abstract{
The equilibrium states associated with the local majority-rule are divided
into three classes, the states of system-wide coordination, the trapped
states, and the states of period-$2$. The effect of clustering coefficient
on the occurrence probability of the states of three classes is analyzed
numerically for Watts-Strogatz and scale-free networks. We further study the
effect of clustering coefficient on the robustness for the states of each
class by proposing a stochastic local majority-rule. The states of period-$2$
are found to be easy to break up, and the trapped states are most robust
among the three classes. For systems in noisy environments, the proposed
stochastic local majority-rule shows that there exists a range of noise for
which, the mean first-passage time from strongly disorder states to the
states of system-wide coordination is the shortest.
} 

\section{Introduction}

\label{intro} The collective behaviors, such as the emergence of system-wide
coordination in nature, the appearance of the consensuses of opinions in
social systems, and other related phenomena, are of great interested for
many researchers\cite{lampl,boyd,krapivsky,young,bikhchandani}. The local
majority-rule (\textit{LMR}) has often been employed to study the arising of
such behaviors. The rule is simple, based on the principle of majority vote
without much consideration in psychological level; it dictates that the time
evolution of the state of an unit (individual or agent) is determined by the
majority-favored state of its neighbors\cite{hopfield}. The neighbors of an
unit can be given by geographic, cultural, social, or organizational
proximity, here we use artificially constructed networks to define the
neighbors of an unit as its nearestly connected nodes. Because of the
locality in \textit{LMR}, one may expect that the distribution of cliques of
a system can affect the occurrence probability of collective behavior, and
this work is devoted to analyze such effect.

The global topology of a network can be characterized by two quantities, the
degree distribution and the clustering coefficient\cite{dorogovtsev}. The
total number of connections of a node is referred as the degree of the node $%
k$, and the probability that a randomly chosen node has $k$ connections is
given by the degree distribution $P\left( k\right) $; the tightness of a
clique formed by a site and its directly connected neighbors can be
characterized globally by the clustering coefficient of a network $C$. The
question concerning with the role of network topology on dynamical
cooperative behavior were discussed by Sood and Render\cite{sood} and by
Suchecki et al.\cite{suchecki} in the voter model which may be viewed as a
statistical model of \textit{LMR}\cite{liggett}. The mean time for reaching
the state of collective behavior was shown to have different scaling
behaviors with respect to the number of nodes for different decay exponents $%
\gamma $ of scale-free (\textit{SF}) networks between $\gamma >3$, $\gamma
=3 $, $2<\gamma <3$, $\gamma =2$, and $\gamma <2$\cite{sood}. Moreover,
network geometry was also shown to have important effect on the dynamics
such as the average survival time of metastable states in finite networks,
the linear size scaling law of the survival time, and the size of an ordered
domain\cite{suchecki}. As \textit{LMR} is the root of the voter model, these
features may be traced to the properties of the equilibrium states of 
\textit{LMR}. Hence the study within \textit{LMR} may provide more insights
to the question.

Different $\gamma $ values of \textit{SF} networks characterize the
difference in the appearance of hub-nodes. Here the hub-nodes are referred
to those possessing large degree of connections. The existence of hub-nodes
may affect strongly the efficiency of reaching an equilibrium state of 
\textit{LMR}. For this aspect, Zhou and Lipowsky showed that there exists
categorical difference between the \textit{SF} networks of $\gamma <5/2$ and
those of $\gamma >5/2$ for the scaling behavior of the relaxation time from
a strongly disorder state towards an order state\cite{zhou}. But there are
other equilibrium states associated with \textit{LMR} that are different
characteristically from the state of collective behavior. Moreover, as a
system evolves from an initial state, the distribution of cliques in the
system affects the corresponding trajectory strongly, and hence affects the
type of equilibrium states reached by the trajectory. One of the questions
we intend to address in this work is as follows: Starting with strongly
disorder states, what role does the clustering coefficient of the system
play in the type of the equilibrium states reached by the system?

Another question that attracts our attention is the relation between the
robustness of a equilibrium state and the clustering coefficient of a
system. The study in this aspect may not only reveal the stability of an
equilibrium state but also provide an estimation for the external strength
required to break the state. The robustness of an equilibrium state can be
characterized by its escape rate after introducing fluctuation to perturb
the system. In fact, fluctuation is an unavoidable component for real
systems. The attempt was made by Moreira et al. to include noise into the
dynamics by changing the transition probability of \textit{LMR} from $1$ to $%
1-\eta $, where the parameter $\eta $ characterizes the average effect of
fluctuation\cite{moreira}. The authors showed that the presence of
fluctuation may increase the probability and the efficiency of occurring
collective behavior for systems with small-world characters. In this work,
we take a microscopic approach by proposing a stochastic \textit{LMR} in
which, each node-state contains a component of white noise. As the
equilibrium states of \textit{LMR} become transient in stochastic dynamics,
this proposal yields the Arrhenius equation for the escape rates. Then, we
determine the dependence of the prefactor and the activation energy of the
Arrhenius equation with respect to the clustering coefficient. This
information allows us to show the effect of the clustering coefficient on
the robustness of a state explicitly. As the state of collective behavior is
of great interested, we also study the mean first-passage time from a
strongly disorder state to the state of collective behavior. Such study, in
addition to the robustness of the state, may provide further understandings
about the role of fluctuation in the process of reaching the state of
collective behavior. \ \ 

This paper is organized as follows. In Sec. II we define the \textit{LMR},
classify its equilibrium states, and briefly describe the generating
processes for the networks used in the numerical study. Based on the \textit{%
LMR}, we numerically calculate the occurrence probabilities of different
classes of equilibrium states for the systems starting with strongly
disorder states, and the results as functions of clustering coefficient are
shown in Sec. III where the dependence on the system size for the occurrence
probability is also discussed. In Sec. IV, we first introduce the stochastic 
\textit{LMR}, then the analysis on the escape rates of different classes of
equilibrium states based on the stochastic dynamics are given. Moreover, the
results for the mean first-passage time to the state of system-wide
coordination are also presented in this section. Finally, a summary of the
results and some general conclusions are given in Sec. V.

\section{Deterministic Dynamics and Networks}

\label{s2}

We first specify the \textit{LMR} and classify the corresponding equilibrium
states. Consider a network system with the distribution of edges given by an 
$N\times N$ adjacency matrix $A$. Here the matrix $A$ is symmetric with the
elements $a_{ij}=1$ for the connected sites $i$ and $j$, and $0$ otherwise.\
The dynamic variable associated with a site $i$ is denoted as $x_{i}$, which
takes two possible values, either $1$\ or $-1$. The system evolves from an
initial to a new configuration in discrete time step according to \textit{LMR%
} whose operation can be either synchronous or asynchronous. In this work,
we consider the synchronous dynamics for which, the rule can be written as 
\begin{equation}
x_{i}\left( t+1\right) ={\mathbf{sgn}}\left( \sum_{j=1}^{N}a_{ij}x_{j}\left(
t\right) \right)  \label{eq001}
\end{equation}%
for $i=1,...,N$, where the \textbf{sgn} function is a standard threshold
function with \textbf{sgn}$\left( x\right) =+1$ for $x>0$ and $-1$ for $x<0$%
, and we set $x_{i}\left( t+1\right) =x_{i}\left( t\right) $ for $%
\sum_{j=1}^{N}a_{ij}x_{j}\left( t\right) =0$. The dynamics of Eq. (\ref%
{eq001}) has been widely studied in discrete neural networks, and the
existence of equilibrium states can be shown by employing the Lyapunov
energy function\cite{hopfield,wilde}, 
\begin{equation}
E\left( t\right) =-\sum_{i=1}^{N}x_{i}\left( t\right) \left[
\sum_{j=1}^{N}a_{ij}x_{j}\left( t-1\right) \right] .  \label{eq002}
\end{equation}%
Moreover, the period of an equilibrium state is either $1$ or $2$\cite{wilde}%
.

The number of equilibrium states indicates the capacity of a neural network.
But, we are interested in the occurrence of the state of collective behavior
and the global characters of other equilibrium states in case that the
collective behavior can not be reached by systems. Thus, the equilibrium
states of period-$1$ are divided into two classes, $S_{0}$ and $S_{1}$. Here
the class $S_{0}$ is for the states of collective behavior for which, all
node-states have the same value, either $1$\ or $-1$; and the class $S_{1}$
consists of all trapped states. For the equilibrium states of period-$2$,
the system oscillates between a pair of states, and we refer all pairs of
equilibrium states as the class $S_{2}$. As the order parameter is defined
as $M_{s}$ $=\sum_{i=1}^{N}x_{i}/N$, the two states of $S_{0}$ have the
value $1$ and $-1$, respectively, and the states of $S_{1}$ and $S_{2}$ have 
$\left\vert M_{s}\right\vert <1$.

Two types of networks, Watts-Strogatz (\textit{WS}) and \textit{SF}
networks, are used to define the adjacency matrix $A$ in the dynamics of Eq.
(\ref{eq001}). The \textit{WS }networks are made from a regular lattice for
which, $N$ sites are placed around a circle and each site has degree, say $%
k_{0}$, connecting to the right and to the left symmetrically, then a
probability $p$ is assigned to rewire the edges randomly\cite{watts}. As the 
$p$ value increases from $0$ to $1$, the resultant network changes from a
regular lattice to a random graph with the clustering coefficient $C$
decreasing from the highest value down to $k_{0}/N$ . Here the $C$ value of
a network is defined as the average of the clustering coefficients
associated with all sites, and the clustering coefficient of a site, say $i$%
, is given as 
\begin{equation}
C_{i}=\frac{2y_{i}}{k_{i}\left( k_{i}-1\right) },  \label{eq002-1}
\end{equation}%
where $y_{i}$ is the number of existent edges between the $k_{i}$ neighbors
of the site $i$.

The \textit{SF} networks are a special category of networks for which, the
degree distributions take the form of power low as $P\left( k\right) \sim
k^{-\gamma }$ with decay exponent $\gamma $. A conventional way of
generating a \textit{SF} network is the scheme of preferential attachment
proposed by Barabasi and Albert\cite{albert1}, and it yields a network with $%
\gamma \approx 2.9$ and small $C$ value $C\approx N^{-0.75}$\cite{albert2}.
The \textit{SF} networks with different $\gamma $ and $C$ values are
employed for numerical study in this work, and they were generated by using
the modified schemes of preferential attachment, proposed by Holme et al.%
\cite{holme} and by Leary et al.\cite{leary}, in a systematic way. For
tuning the $C$ value without altering the $\gamma $ value, a step, called
triad-formation, is added to the process of preferential attachment with a
assigned probability; the larger the assigned probability of performing
triad-formation, the larger the $C$ value of the resultant network is\cite%
{holme}. Alternatively, we alter the $\gamma $ value without affecting the $%
C $ value by switching the uniform distribution for the random numbers used
in the process of preferential attachment to the distribution of a designed
probability density function\cite{leary}. As the designed probability
density function further enhances the probability of connecting a new edge
to a site with large degree, we obtain a network with a larger $\gamma $
value. On the other hand, for the opposite tendency in the designed function
the resultant network has a smaller $\gamma $ value.

\section{Occurrence Probabilities of Equilibrium States}

\label{s3} \ \ \ 

We first calculate numerically the occurrence probability for three
different classes of equilibrium states of Eq. (\ref{eq001}) to analyze the
clustering effect. The occurrence probabilities denoted as $P_{i}$ for the
class $S_{i}$ of the equilibrium states with $i=0$, $1$, or $2$ for systems
modelled by the \textit{WS} and the \textit{SF} networks. To minimize
statistical errors in the simulation results, we generate $1000$ samples for
the \textit{WS }networks with a given value of rewiring probability, and the
corresponding clustering coefficient $\overline{C}$ is defined as the
average of the coefficients of all samples. Then, the $P_{i}$ value
associated with a $\overline{C}$ is given by the fractional percentage of
occurrence for the equilibrium states belonging to the class $S_{i}$ over
the $10^{6}$ trajectories. Here the trajectories all start from strongly
disorder states and are equally distributed in the $1000$ samples.

The results of $P_{i}$ for the \textit{WS} networks are shown as the plots
of $P_{i}$ vs. $\overline{C}$ in Fig. 1(a)\ for different $k_{0}$ values
with $N=1000$ and in Fig. 1(b) for different $N$ values with $k_{0}=8$. \
Here the $\overline{C}$ values are in the range $0<\overline{C}<0.5$ for
which, the corresponding rewiring probabilities are ranged between $%
0.04<p<0.5$; this is also the range of $\overline{C}$ generated for the 
\textit{SF} networks. As shown in the plots, systems are more likely to be
led to the states of collective behavior for small $\overline{C}$ values $%
\left( \overline{C}\sim 0.05\right) $, to the equilibrium states of $S_{1}$
for medium $\overline{C}$ values $\left( \overline{C}\sim 0.15\right) $, and
to the equilibrium states of $S_{2}$ for high $\overline{C}$ values $\left( 
\overline{C}\sim 0.3\right) $. The results also indicate that the $P_{0}$
value is suppressed and the $P_{2}$ value is enhanced as the node number $N$
increases, and the tendency is opposite for increasing the average number of
degree $k_{0}$.

Analytic expressions $P_{i}\left( N,\overline{C}\right) $, which can fit the
data shown in Fig. 1(b) properly, are found for a better understanding about
the $N$-dependence of the occurrence probabilities $P_{i}$. The results are 
\begin{equation}
P_{0}\left( N,\overline{C}\right) =\frac{1.11-\left( 0.06\right) N^{0.16}}{%
1+\exp \left\{ \left[ 91.53-\left( 407.28\right) /N^{0.35}\right] \overline{C%
}-6.70\right\} },  \label{eq002-2}
\end{equation}%
\begin{equation}
P_{2}\left( N,\overline{C}\right) =1.03-\frac{0.95+\left( 1.04\times
10^{5}/N^{2.07}\right) z\left( N,\overline{C}\right) }{1+z\left( N,\overline{%
C}\right) },  \label{eq002-3}
\end{equation}%
and 
\begin{equation}
P_{1}\left( N,\overline{C}\right) =1-P_{0}\left( N,\overline{C}\right)
-P_{2}\left( N,\overline{C}\right) ,  \label{eq002-4}
\end{equation}%
where $z\left( N,\overline{C}\right) $ is 
\begin{equation}
z\left( N,\overline{C}\right) =\left[ \frac{\overline{C}}{\left(
0.65/N^{0.07}-0.25\right) }\right] ^{5}.  \label{eq002-5}
\end{equation}%
As shown as the solid lines in Fig. 1(b), the expressions agree with the
numerical data very well. Eq. (\ref{eq002-2}) reveals an important feature
about $P_{0}$, that is, the $P_{0}$ value is suppressed as increasing $N$ or 
$\overline{C}$. Consequently, there exists a site number $N^{\ast }\left(
k_{0}\right) $ that $P_{0}\approx 0$ for $N>N^{\ast }\left( k_{0}\right) $,
where $N^{\ast }\left( k_{0}\right) $ decreases as $k_{0}$ decreases.
Moreover, Eq. (\ref{eq002-3}) indicates that the $P_{2}$ value increases as $%
N$ increases. Thus, we may have $P_{0}\approx 0$, $P_{1}\approx 0$, and $%
P_{2}\approx 1$ for $N>N^{\ast }\left( k_{0}\right) $. This is demonstrated
by the numerical results shown in Fig. 2 where the $P_{i}$ values as
functions of $\overline{C}$ for $N=6000$ and $k_{0}=4$ are given.

The statistics for the numerical study in the \textit{SF} networks is
similar to the case of the \textit{WS }networks. As the modified schemes of
preferential attachment are applied, the $\gamma $ value may change slightly
in tuning the $C$ value\cite{holme}, and the $C$ value may alter slightly in
tuning the $\gamma $ value\cite{leary}. Thus, the samples are characterized
by the set-values $\left( \overline{\gamma },\overline{C}\right) $ with $%
1000 $ members belonging to a $\left( \overline{\gamma },\overline{C}\right) 
$. There are three distinct $\overline{\gamma }$ values, $\overline{\gamma }%
=2.82$, $2.55$, and $2.29$, and the possible $\overline{C}$ values for a $%
\overline{\gamma }$ value are in the range between $0$ and $0.5$. All
samples have the site number $N=5000$ and the average degree of a site $%
\left\langle k\right\rangle =4$. The $P_{i}$ value is the result over the $%
10^{6}$ trajectories for which, they are equally distributed in the $1000$
samples of the \textit{SF} networks belonging to the same $\left( \overline{%
\gamma },\overline{C}\right) $ and each trajectory starts from a strongly
disorder state. The numerical results are shown as the plots of $P_{i}$ vs. $%
\overline{C}$ in Fig. 3 for three different $\overline{\gamma }$ values.

The results obtained from the \textit{SF} networks indicate that although
the $\overline{\gamma }$ value may affect significantly the convergent speed
of the system leading to a equilibrium state, it has little effect on the $%
P_{i}$ value for which, the $\overline{C}$ value plays a major role.
Moreover, similar to the case of the \textit{WS} networks, as shown in Fig.
3 the clustering coefficient $\overline{C}$ drives the system from the state
of collective behavior at low $\overline{C}$ $\left( \overline{C}\leq
0.1\right) $\ to the the phase of oscillation between two states at high $%
\overline{C{\ }}$ $\left( 0.5>\overline{C}>0.3\right) $, and the system is
trapped in a state of $S_{1}$ at medium $\overline{C}$. One may further
expect that the $N$-dependence for $P_{i}$ has the same feature
qualitatively as that for the \textit{WS} networks.

\section{Stochastic Dynamics}

\label{s4} \ \ 

Noise is a very natural component for real systems, its physical origins can
be traced to incomplete information, processing errors, or other
environmental perturbations. To add a component of noises to the \textit{LMR}%
, we propose a stochastic version based on the assumption that the effect of
noise is localized and appears as the fluctuation in recognizing the value
of a node-state by its connected neighbors. Then, the stochastic \textit{LMR}
is given as 
\begin{equation}
x_{i}\left( t+1\right) =\mathbf{sgn}\left( \sum_{j=1}^{N}a_{ij}\mathbf{sgn}%
\left( x_{j}\left( t\right) +\sqrt{2D}\xi _{j}\left( t\right) \right)
\right) ,  \label{eq05}
\end{equation}%
where $\xi _{i}\left( t\right) $ is the Gaussian white noise with the zero
mean and the $\delta $-function correlation, i,e, $\left\langle \xi
_{i}\left( t\right) \right\rangle =0$ and $\left\langle \xi _{i}\left(
t\right) \xi _{j}\left( s\right) \right\rangle =\delta _{i,j}\delta \left(
t-s\right) $, and $D$ is the diffusion constant which characterizes the
strength of noise. The equilibrium states of Eq. (\ref{eq001}) become
transient for the stochastic dynamics of Eq. (\ref{eq05}). Consequently, the
escape time for different classes of the equilibrium states can be measured,
the results are denoted as $\left\langle \tau ^{\left( i\right)
}\right\rangle $ with $i=0$, $1$, and $2$ for the class $S_{0}$, $S_{1}$,
and $S_{2}$, respectively, and the inverse of $\left\langle \tau ^{\left(
i\right) }\right\rangle $ yields the escape rate $\left\langle \kappa
_{i}\right\rangle $, where the bracket of $\tau ^{\left( i\right) }$ or $%
\kappa _{i}$ represents the average of the results over the samples of
different networks.

As the stochastic dynamics of Eq. (\ref{eq05}) is applied to the equilibrium
states of Eq. (\ref{eq001}), one can expect that the escape rates obey the
Arrhenius equation\cite{kampen}, 
\begin{equation}
\left\langle \kappa _{i}\right\rangle =A_{i}\exp \left( -\frac{\Delta G_{i}}{%
D}\right) ,  \label{eq06}
\end{equation}%
where $A_{i}$ is the prefactor, and $\Delta G_{i}$ is the activation energy.
Note that based on the fluctuation-dissipation theorem we may identify $%
D=k_{B}T$ with the Boltzman constant $k_{B}$ and the absolute temperature $T$%
\cite{kubo}. The factor $A_{i}$, which is equivalent to the rate constant in
a chemical reaction, signifies the entropy effect, and one can expect that
it has a strong dependence on the clustering coefficient. The value of $%
\Delta G_{i}$ gives the maximum potential energy required to escape from the
equilibrium state. Thus, the results of $A_{i}$ and $\Delta G_{i}$ provide
insights on the robustness of the equilibrium states of different classes as
the geometric structure of a system varies. \ 

We perform the numerical measurements for $\left\langle \kappa
_{i}\right\rangle $ in the \textit{SF} networks, and the statistics for the
measurements is described as follows. We first single out the $m_{0}$
equilibrium states belonging to the class $S_{i}$ for a set-value $\left( 
\overline{\gamma },\overline{C}\right) $ with $m_{0}=25000$, where the $%
m_{0} $ states of the class $S_{0}$ are identical as $x_{i}=+1$ for all
nodes. Then, the number of time-steps required to escape from each of the $%
m_{0}$ states is measured with a preassigned cut-off time-step $t_{\max }$,
such a measurement is repeated for $20$ times to realize the Gaussian
distribution of the noise $\xi _{i}\left( t\right) $, and the escape
time-step $\left\langle \tau ^{\left( i\right) }\right\rangle $ for the
class $S_{i}$ is given by the average value over the $20$ simulations and
over the $m_{0}$ states. \ 

Our results for the \textit{SF} networks with $\left( \overline{\gamma },%
\overline{C}\right) =\left( 2.82,0.3026\right) $ are shown in Fig. 3 as the
plots of $\log \left\langle \tau ^{\left( i\right) }\right\rangle $ vs. $1/D$
with $i=0$, $1$, and $2$, and those for different $\left( \overline{\gamma },%
\overline{C}\right) $ values also yield straight lines in the same plots.
The results indicate that the escape rates $\left\langle \kappa
_{i}\right\rangle $ for the equilibrium states belonging to the class $S_{i}$
agree with Eq. (\ref{eq06}), such agreement persists as the cut-off
time-steps $t_{\max }$ increases from $10000$ to $20000$ as shown in Fig. 4.
Then, the activation energy $\Delta G_{i}$ of Eq. (\ref{eq06}) is determined
from the slope of $\log \left\langle \tau ^{\left( i\right) }\right\rangle $
vs. $1/D$, it yields $\Delta G_{0}=0.81$ and $\Delta G_{1}=\Delta G_{2}=1.0$%
, independent of the $\left( \overline{\gamma },\overline{C}\right) $ value.
Moreover, the prefactor $A_{i}$ can be determined by the intersection
between the straight line of $\log \left\langle \tau ^{\left( i\right)
}\right\rangle $ vs. $1/D$ and the vertical line of $1/D=0$, we have $%
A_{0}=97.08$, which is independent of the $\left( \overline{\gamma },%
\overline{C}\right) $ value, and the other two prefactors $A_{1}$ and $A_{2}$
depend on the network geometry. The $\overline{\gamma }$-dependence may
occur for $A_{S_{1}}$ when $\overline{C}$ is large, for example, we have $%
A_{S_{1}}=10.11$, $7.20$, and $5.88$ for $\overline{C}=0.3$ but $\overline{%
\gamma }=2.29$, $2.55$, and $2.82$, respectively; such dependence is not
found for $A_{S_{2}}$. On the other hand, both $A_{1}$ and $A_{2}$ are very
sensitive to the $\overline{C}$ value, and their $\overline{C}$-dependence
is shown explicitly in Fig. 5 for $\overline{\gamma }=2.82$. \ \ 

Some important features for Eq. (\ref{eq06}) are revealed from the results
shown in Figs. 4 and 5. Firstly, the decay rate for the state of collective
behavior is universal in the sense that it is independent of the network
geometry, so are the $\Delta G_{i}$ values for the equilibrium states
belonging to the classes $S_{1}$ and $S_{2}$. Moreover, the activation
energy for the states of $S_{1}$ and $S_{2}$ is larger than that for the
states of $S_{0}$. For the entropy effect on the decay rate, the $\overline{C%
}$ value has a significant role in determining the values of $A_{1}$ and $%
A_{2}$. As the results of Fig. 5 indicate, the $A_{2}$ value decreases for
increasing the $\overline{C}$ value, and the $A_{1}$ value behaves
oppositely; moreover, the equilibrium states of $S_{2}$ possess a very large
prefactor, and this overcomes the higher activation energy and renders the
states to be very fragile. Then, as the conceptual sketch shown in Fig. 6
for the potential barriers of the equilibrium states belonging to different
classes, we may conclude that the states of $S_{2}$ are very easy to\ break
up, and the states of $S_{1}$ are the most robust among the equilibrium
states of three classes.

As the state of collective behavior is of great interested, we then study
its mean first-passage time for the system with the \textit{SF} networks in
a noisy environment. In the numerical calculations, we first assign a
strongly disorder configuration to the system, then follow the stochastic 
\textit{LMR} of Eq. (\ref{eq05}) to generate a trajectory and to record the
time-steps for the first appearance of a state of $S_{0}$. Here, the
recorded time-steps is set as $t_{\max }=10^{4}$ for the absence of the
states of $S_{0}$ at $t=t_{\max }$. We generate $10^{6}$ trajectories
totally for a given set of $\left( \overline{\gamma },\overline{C}\right) $,
and the mean first-passage time, denoted as $\left\langle \tau
_{0}\right\rangle $, is given\textit{\ }as the average of the recorded
time-steps over all trajectories. As the differences in $\left\langle \tau
_{0}\right\rangle $ caused by different $\gamma $ values are insignificant,
we show the results as the plot of $\left\langle \tau _{0}\right\rangle $
vs. $D$ for $\overline{\gamma }=2.82$ and $\overline{C}=0.0066$, $0.2016$,
and $0.3026$, respectively in Fig. 7. It is interesting to observe that the $%
\left\langle \tau _{0}\right\rangle $ value as a function of $D$ is
non-monotonic. As the $D$ value increases, the $\left\langle \tau
_{0}\right\rangle $ value first decreases, reaches a minimum flat valley in
the range of $0.07\lesssim D\lesssim 0.22$ with $\left\langle \tau
_{0}\right\rangle <10$, then increases abruptly at $D_{\max }\simeq 0.22$,
and the state of collective behavior becomes unreachable for $D>D_{\max }$.
Moreover, the role of clustering coefficient in the mean first-passage time
is not very significant, as the results shown in Fig. 5 indicate, the
difference in clustering coefficients is noticed by the mean first-passage
time for $D\leq 0.10$.

\section{Summary and Conclusion}

\label{s5} \ \ 

In summary, we show explicitly how the clustering coefficient of a system
affect the occurrence probability of the different types of equilibrium
states associated with the \textit{LMR}. As the state of system-wide
coordination is concerned, the increasing of the clustering coefficient
would suppress its probability of occurrence. On the other hand, systems
with large clustering coefficients are easily led to trapped states or
oscillations between pairs of states. We also propose a stochastic version
of the \textit{LMR} for which, the decay rate of an equilibrium state obeys
the Arrhenius equation. This allows us to quantify the robustness of the
equilibrium states through the values of the prefactor and the activation
energy in the Arrhenius equation. Our results indicate that the states of
period-$2$ are very fragile, and the trapped states are the most robust
among the three types. For systems in noisy environments, our results
obtained from the stochastic \textit{LMR} indicate that there exists a range
of noise for which, the mean first-passage time from strongly disorder
states to the states of system-wide coordination is the shortest. Thus, the
efficiency in reaching the state of collective behavior may be improved for
systems with certain amount of noise. As the distribution of subgroups in a
social system can be characterized by the clustering coefficient, all these
results may provide wide applications in the study of collective behaviors
of social systems.

\textbf{Acknowledgement}: We would like to thank Dr. Yutin Huang for
fruitful discussions and carefully reading and editing the manuscript. We
also thank the National Center for High-performance Computing for providing
the computing facilities.This work was partially supported by the National
Science Council of Republic of China (Taiwan) under the Grant No. NSC
96-2112-M-033-006 (MCH) and 99-2112-M-150-002 (YPL).

\end{document}